\documentstyle[amsfonts,prb,aps,preprint,epsfig]{revtex}  
  
\ifpreprintsty        
\makeatletter       
\def\@CITEX[#1]#2{\if@filesw\immediate\write\@auxout{\string\citation{#2}}\fi       
\leavevmode [\@cite{\@collapse{#2}}{#1}]}   
\makeatother       
\fi       
   
\newcommand{\Sqi}{\smash{\hat{\vec{S}_i}}{\vphantom{\hat{S}}}^2}   
\newcommand{\qq}{\smash{\vec{q}}{\vphantom{q}\,}^2}   
\begin{document}   
\draft  
  
\title{Random Phase Approximation for Multi-Band Hubbard Models}   
  
\author{J\"org B\"unemann}  
\address{Oxford University, Physical and Theoretical Chemistry Laboratory,   
South Parks Road, Oxford OX1 3QZ, UK}  
  
\author{Florian Gebhard}  
\address{Fachbereich Physik, Philipps-Universit\"at Marburg, D-35032   
Marburg, Germany}

\date{{\bf Preprint version of July 1, 2001}}  
\maketitle   
   
\begin{abstract}%  
We derive the random-phase approximation for spin excitations   
in general multi-band Hubbard models,    
starting from a collinear ferromagnetic Hartree-Fock ground state. 
The results are compared with those of a   
recently introduced variational many-body approach  
to spin-waves in itinerant ferromagnets.  
As we exemplify for Hubbard models with one and two bands,  
the two approaches lead to qualitatively different results.  
The discrepancies can be traced back to the fact  
that the Hartree-Fock theory fails to describe properly  
the local moments which naturally arise in a correlated-electron  
theory.  
\end{abstract}   
   
  
\section{Introduction}   
\label{int}   
   
Metallic ferromagnetism is one of the oldest problems in solid-state physics.   
Despite all progress in solid-state theory over the last  
century~\cite{stoner,vanVleck,gutzi,Herring,moriya,Capellmann,yosida,Fazekas},   
we are still far away from a definite theory for   
transition metals like iron, cobalt and nickel.  
It is generally accepted that the magnetic order in metals is a consequence  
of the interplay between the electrons' kinetic energy and their mutual  
Coulomb interactions. Therefore, it does not come as a surprise  
that a complete theory for this many-particle effect is difficult to develop.  
  
Nevertheless, it is commonly   
believed that at least the ground-state properties of itinerant magnets  
can be described by (effective) one-particle theories   
such as Hartree-Fock theory or (spin-)density-functional theory (SDFT).   
However, serious doubts are in order,   
as a closer look on available experimental  
data reveals. For example, SDFT fails  
to reproduce even qualitatively the electronic band-structure of   
nickel, one of the simplest itinerant ferromagnets~\cite{Eberhardt};  
for a summary, see Ref.~\onlinecite{gutzjubel}.   
For iron, the agreement is better but   
cannot be considered as satisfactory either.  
In any case, the same physical mechanism is acting in both systems,  
and a convincing theory should cover all elements of the iron group  
equally well.  
  
Single-particle theories are not very satisfactory also from a purely  
theoretical point of view: It is well known that they  
significantly overestimate the stability of ferromagnetism~\cite{stoner}.  
For example, Hartree-Fock theory for one-band Hubbard models  
gives a ferromagnetic ground state above some   
moderate interaction strength. In contrast,  
many-body theories show that ferromagnetism in  
one-band Hubbard models occurs, if at all, only for very large  
Coulomb interactions or very peculiar choices of the  
one-particle density of states~\cite{Vollhardtrev,ulmke}.  
This indicates that the Stoner mechanism is less prominent  
in itinerant ferromagnets than implied by   
single-particle theories.  
  
As already argued by van Vleck~\cite{vanVleck} and confirmed  
recently~\cite{PRB98,vollhardt,Reinhard,Japaner},  
an essential ingredient for metallic ferromagnetism is the local  
(Hund's-rule) exchange coupling of electrons, a generic feature  
of electrons in degenerate bands.   
In contrast to the one-band case, ferromagnetism  
occurs for moderate Coulomb-interactions in a generic two-band model  
provided that local Hund's-rule terms  
are included.  
Therefore, even a qualitative understanding of the 3$d$-elements  
requires a careful consideration of their atomic open-shell structures.  
  
In our recent work~\cite{PRB98,JPhys} we   
have developed a variational approach which allows us to   
examine multi-band Hubbard models with an arbitrary  
number of correlated orbitals per lattice site.  
For a realistic description of the  
3$d$-elements a minimal model includes the 3$d$, 4$s$, and 4$p$  
bands. When applied to nickel, our theory provides results which agree  
much better with experiments than all previous   
SDFT-calculation~\cite{gutzjubel,wandlitz}. In particular,  
our approach reproduces the experimental Fermi-surface topology of nickel  
and removes all other qualitative SDFT deficiencies.  
Since our variational many-body approach provides a reasonable  
description of the ground state, it may serve as  
an appropriate starting point for the study of excitations  
in itinerant ferromagnets.  
  
As observed in inelastic neutron scattering experiments~\cite{lovesey,mook},  
the spectrum of itinerant ferromagnets displays low-lying  
spin-wave excitations.   
These excitations are generally considered to be the driving force  
for the transition from the ferromagnetic phase into the paramagnetic  
phase at the Curie temperature. Therefore,  
their correct description is also very important for our   
understanding of finite-temperature properties of itinerant ferromagnets.  
Unfortunately, almost all established theories which are used for  
the analysis of spin-waves in transition metals are based on  
effective single-particle theories which do not   
necessarily provide a satisfactory description of the ground state.  
  
The textbook approach to the problem of itinerant ferromagnetism is a  
Hartree-Fock theory for the ground state which is then combined with a  
random-phase approximation (RPA) for the description of   
spin-excitations~\cite{moriya,yosida}.   
For example, in Ref.~\onlinecite{cook}   
the RPA-method has been applied to iron and nickel  
starting from a SDFT calculation for the ground state.  
Later, a similar method has been used to analyze   
data from electron energy-loss spectroscopy for the same materials~\cite{mills}.  
Another common approach starts from energy calculations  
for static spin deformations of the ferromagnetic ground state within the  
SDFT. The spin-wave energies are identified as the energy difference between  
these ``frozen-magnon'' states and the ground state\cite{eschrig}.  
Despite their conceptual shortcomings both SDFT-based   
approaches are able to reproduce the experimental data surprisingly well.  
  
Very recently~\cite{spinpaper} one of us has proposed a spin-wave   
theory which starts from  
the variational ground state of multi-band Hubbard models as   
introduced in Ref.~\onlinecite{PRB98}.   
This theory allows us to determine the spin-wave  
spectrum for systems with an arbitrary number of correlated orbitals per  
lattice site. The main advantage of this new approach lies in its  
foundation on a true many-particle description of the ground  
state. In this way, we are able to provide  
a consistent picture of the ferromagnetic Fermi liquid and   
its spin-wave excitations.   
  
In order to obtain a better understanding of our own   
but likewise of the SDFT-based methods it is crucial to   
apply them for a generic but simple model system.  
To our knowledge such an analysis has  
not been performed yet. Even a comparison between the  
two well-established SDFT methods is still lacking.  
In this work we analyze the Hartree-Fock/RPA theory  
and compare it with our variational many-body approach.  
Although the RPA is generally considered as the ``standard''  
method for the description of magnetic   
excitations in itinerant ferromagnets~\onlinecite{moriya}    
there exists no systematic derivation for multi-band Hubbard models.    
In Sect.~\ref{rand}, we derive the general RPA equations   
from a decoupling scheme and a standard diagrammatic viewpoint.  
  
In Sect.~\ref{comp} we provide numerical results for a generic two-band   
model and show that both methods lead to  
qualitatively different results. For generic values of the magnetization,  
the Hartree-Fock/RPA treatment incorrectly predicts  
an instability of the collinear ferromagnetic phase against some  
spiral ordering.   
In our variational approach the spin-waves always have positive energy   
and their stiffness decreases with increasing correlations.  
The case of a one-band model with fully polarized bands  
allows us to pinpoint the origin of these discrepancies:  
The Hartree-Fock theory allows for ferromagnetism with  
small local moments and  
large longitudinal spin-fluctuations due to Stoner excitations.  
These are actually absent in a realistic correlated-electron description  
where charge fluctuations are strongly suppressed in the   
ferromagnetic and paramagnetic phases.  
  
Conclusions in Sect.~\ref{summ} close our presentation.  
  
\section{Random phase approximation for multi-band systems}   
\label{rand}   
   
In this section we derive the random phase approximation (RPA) for   
ferromagnetism in multi-band Hubbard models.    
Starting from a LDA-calculation, RPA results for iron and   
nickel have been reported in Ref.~\onlinecite{cook}. However, in these   
calculations the relevant Coulomb matrix elements were significantly   
simplified. In this work we present a more general treatment.   
   
\subsection{General remarks}   
\label{gene}   
   
We will address the following general class of multi-band Hubbard   
models~\cite{PRB98,florianbuch}    
\begin{equation}   
\hat{H}=\sum_{i,j}\sum_{s,s^{\prime }}t_{i,j}^{s,s^{\prime }}   
\hat{c}_{i;s}^{+}\hat{c}_{j;s^{\prime }}^{\vphantom{+}}   
+\sum_{i}\hat{H}_{i;\text{at}}\equiv \hat{H}_{1}+\hat{H}_{\text{at}}\;.   
\label{1.1}   
\end{equation}   
Here, $\hat{c}_{i;s}^{+}$ creates an electron with combined spin-orbit   
index~$s=(b,\sigma)$ at the lattice site~$\vec{R}_i$   
of a solid ($b=1,\ldots,N$; $\sigma=\uparrow,\downarrow$;   
$N=5$ for 3$d$~electrons).   
The atomic Hamiltonian    
\begin{equation}   
\hat{H}_{i;\text{at}}=\sum_{s_{1},s_{2},s_{3},s_{4}}   
{\cal U}^{s_{1},s_{2};s_{3},s_{4}}   
\hat{c}_{i;s_{1}}^{+}\hat{c}_{i;s_{2}}^{+}\hat{c}_{i;s_{3}}^{\vphantom{+}}   
\hat{c}_{i;s_{4}}^{\vphantom{+}}  \label{1.2}   
\end{equation}   
is assumed to have site-independent interaction parameters   
${\cal U}^{s_{1},s_{2};s_{3},s_{4}}$.   
   
In order to determine the spin-wave properties, we need   
to consider the imaginary part $\chi_{T}(\vec{q},E)$ of the transverse   
susceptibility\cite{lovesey}, which is given by the retarded two-particle   
Green function    
\begin{mathletters}   
\label{1.3}   
\begin{eqnarray}   
G_{\text{ret}}(\vec{q},E)    
&=&\frac{1}{L}\langle \langle    
\hat{S}_{\vec{q}}^{+}\,;\hat{S}_{\vec{q}}^{-}\rangle\rangle_{E}     
\label{1.3a} \\   
&=&-\frac{i}{L}\int_{0}^{\infty }dt\,e^{iEt}   
\left\langle \Psi_{0}\left| \left[    
\hat{S}_{\vec{q}}^{+}(t),\hat{S}_{\vec{q}}^{-}(0)   
\right] \right|    
\Psi_{0}\right\rangle \;.  \label{1.3b}   
\end{eqnarray}\end{mathletters}%   
Here, we introduced the $\vec{q}$-dependent spin-flip operators    
\begin{mathletters}   
\label{1.4}   
\begin{eqnarray}   
\hat{S}_{\vec{q}}^{+} &=&\sum_{l}e^{i\vec{q}\vec{R}_{l}}   
\hat{S}_{l}^{+}=\sum_{l,b}e^{i\vec{q}\vec{R}_{l}}   
\hat{c}_{l;(b\uparrow )}^{+}\hat{c}_{l;(b\downarrow )}^{{}}\;,     
\label{1.4a} \\   
\hat{S}_{\vec{q}}^{-} &=&(\hat{S}_{\vec{q}}^{+})^{+}=   
\sum_{l,b}e^{-i\vec{q}\vec{R}_{l}}   
\hat{c}_{l;(b\downarrow )}^{+}\hat{c}_{l;(b\uparrow )}^{{}} \; ,  
\label{1.4b}   
\end{eqnarray}\end{mathletters}%   
in the Heisenberg picture, where the sum runs over all $L$ lattice sites    
$\vec{R}_l$ and orbitals $b$.   
   
Magnetic excitations are found as poles of the Green function    
$G_{\text{ret}}(\vec{q},E)$, or, equivalently,    
as peaks in    
\begin{equation}   
\chi_{T}(\vec{q},E)= \mathop{\rm Im} G_{\text{ret}}(\vec{q},E)   
\end{equation}    
at energies $E>0$. The Lehmann representation of~(\ref{1.3}),    
\begin{equation}   
G_{\text{ret}}(\vec{q},E)=   
\frac{1}{L}\sum_{n}\left[    
\frac{\mid \langle \Psi _{n}\mid    
\hat{S}_{\vec{q}}^{-}   
\mid \Psi _{0}\rangle \mid ^{2}}{E-(E_{n}-E_{0})+i\eta}   
-\frac{\mid \langle \Psi _{n}\mid    
\hat{S}_{\vec{q}}^{+}\mid \Psi_{0}\rangle \mid ^{2}}{E+(E_{n}-E_{0})+i\eta }   
\right] \; ,  \label{1.5}   
\end{equation}   
shows that the weights of the poles in $G_{\text{ret}}(\vec{q},E)$    
at the energies $E_{n}-E_{0}>0$   
are given by   
\begin{equation}   
W_n=\mid \langle \Psi _{n}\mid    
\hat{S}_{\vec{q}}^{-}\mid \Psi_{0}\rangle \mid ^{2} \; .   
\end{equation}   
In a ferromagnetic system the state    
\begin{equation}   
|\Psi _{\vec{q}}^{0}\rangle \equiv    
\hat{S}_{\vec{q}}^{-}|\Psi _{0}\,\rangle \label{swf}   
\end{equation}   
is also a ground state of $\hat{H}$ for $\vec{q}=\vec{0}$ because the   
operator $\hat{S}_{\vec{q}=\vec{0}}^{-}$ just flips a spin in the   
spin multiplet of the ground state $\left|\Psi_{0}\right\rangle $.   
Therefore, we can conclude that $G_{T}(\vec{0},E)$ has one isolated pole for    
$E-E_{0}=0$. This is not surprising since the spin-wave is the gapless   
Goldstone mode in the symmetry-broken ferromagnetic phase. For finite values   
of $\vec{q}$, it is an experimental fact that there are also pronounced   
peaks in $\chi_{T}(\vec{q},E)$ at the spin-wave energies    
\begin{equation}   
E=E(\vec{q}\,) \approx D\qq \; .  \label{1.6}   
\end{equation}   
The constant $D$ is usually denoted as the ``spin-wave stiffness''.   
   
\subsection{Hartree-Fock treatment}   
\label{hart}   
   
The one-particle Hamilton $\hat{H}_{1}$ in~(\ref{1.1}) may be written in   
momentum space as    
\begin{mathletters}   
\label{2.4}   
\begin{eqnarray}   
\hat{H}_{1} &=&\sum_{\vec{k};b,b^{\prime },\sigma }   
\varepsilon_{b,b^{\prime}}(\vec{k})   
\hat{c}_{\vec{k};(b\sigma )}^{+}\hat{c}_{\vec{k};(b^{\prime}\sigma )}^{{}}     
\label{7.1} \\   
&=&\sum_{\vec{k};\alpha,\alpha^{\prime },\sigma }   
E_{\alpha \alpha ^{\prime}}(\vec{k},\sigma )\,   
\hat{a}_{\vec{k};\alpha ,\sigma }^{+}   
\hat{a}_{\vec{k};\alpha ^{\prime },\sigma }\;.   
\end{eqnarray}\end{mathletters}%   
Here, we introduced the operators    
\begin{mathletters}   
\label{7.2}   
\begin{eqnarray}   
\hat{a}_{\vec{k};\alpha ,\sigma }^{+} &=&   
\sum_{b=1}^N u_{\alpha ,b}^{{}}(\vec{k},\sigma )   
\hat{c}_{\vec{k};(b\sigma )}^{+}\;,  \label{7.2a} \\   
\hat{a}_{\vec{k};\alpha ,\sigma }^{{}} &=&   
\sum_{b=1}^N u_{\alpha ,b}^{\ast }(\vec{k},\sigma )   
\hat{c}_{\vec{k};(b\sigma )}^{{}}\;,  \label{7.2b}   
\end{eqnarray}\end{mathletters}%   
and the respective energy matrix-elements    
\begin{equation}   
E_{\alpha \alpha ^{\prime }}(\vec{k},\sigma )=\sum_{b,b^{\prime}}   
\varepsilon_{b,b^{\prime }}(\vec{k})u_{\alpha ,b}^{\ast }(\vec{k},\sigma)   
u_{\alpha ^{\prime },b^{\prime }}^{{}}(\vec{k},\sigma )\;,   
\end{equation}   
where the elements $u_{\alpha ,b}(\vec{k},\sigma )$ of the unitary matrix    
$u(\vec{k},\sigma )$ will be identified later.    
The transformation~(\ref{7.2})   
leads to the following expression for the atomic Hamiltonian,    
\begin{equation}   
\hat{H}_{\text{at}}=\frac{1}{L}   
\sum_{\vec{p},\vec{p}\,^{\prime },\vec{q}\,^{\prime }}   
\sum_{\beta _{1},\beta _{2},\beta _{3},\beta _{4},\sigma,\sigma^{\prime }}   
V_{\beta _{1}\beta _{4}\sigma }^{\beta _{2}\beta_{3}\sigma ^{\prime }}%   
(\vec{p},\vec{p}\,^{\prime }+\vec{q}\,^{\prime },\vec{p}\,^{\prime },   
\vec{p}+\vec{q}\,^{\prime })   
\hat{a}_{\vec{p};\beta_{1},\sigma}^{+}   
\hat{a}_{\vec{p}\,^{\prime }+\vec{q}\,^{\prime };\beta _{2},\sigma^{\prime }}^{+}   
\hat{a}_{\vec{p}\,^{\prime };\beta_{3},\sigma ^{\prime}}^{{}}   
\hat{a}_{\vec{p}+\vec{q}\,^{\prime };\beta _{4},\sigma }^{{}}\;,   
\label{7.3}   
\end{equation}   
where    
\begin{mathletters}   
\label{7.4}   
\begin{eqnarray}   
V_{\beta _{1}\beta _{4}\sigma }^{\beta _{2}\beta _{3}\sigma ^{\prime }}   
(\vec{p}_{1},\vec{p}_{2},\vec{p}_{3},\vec{p}_{4})   
&=&   
\sum_{b_{1},b_{2},b_{3},b_{4}}   
U_{\sigma ,\sigma ^{\prime}}^{b_{1},b_{2},b_{3},b_{4}}   
u_{\beta _{1},b_{1}}^{\ast }(\vec{p}_{1},\sigma)   
u_{\beta _{2},b_{2}}^{\ast }(\vec{p}_{2},\sigma ^{\prime })   
u_{\beta_{3},b_{3}}^{{}}(\vec{p}_{3},\sigma ^{\prime })   
u_{\beta _{4},b_{4}}^{{}}(\vec{p}_{4},\sigma )\;, \nonumber\\   
&&  \label{7.4a} \\   
U_{\sigma ,\sigma ^{\prime }}^{b_{1},b_{2},b_{3},b_{4}}    
&\equiv &   
{\cal U}^{(b_{1}\sigma),(b_{2}\sigma^{\prime});   
(b_{3}\sigma^{\prime}),(b_{4}\sigma )}\;.   
\end{eqnarray}\end{mathletters}%   
We choose the matrix $u(\vec{k},\sigma )$ in order to fulfill the   
corresponding Hartree-Fock equations    
\begin{eqnarray}   
E^{\text{HF}}_{\beta }(\vec{k},\sigma )\delta _{\beta \beta ^{\prime }}   
&=&   
E_{\beta \beta ^{\prime }}(\vec{k},\sigma)   
+2\sum_{b_{1},b_{2},b_{3},b_{4},\sigma ^{\prime }}   
U_{\sigma ^{\prime},\sigma }^{b_{1},b_{2},b_{3},b_{4}}   
\left[ T_{b_{1},b_{4}}^{\sigma ^{\prime}}   
u_{\beta ,b_{2}}^{\ast }(\vec{k},\sigma )   
u_{\beta ^{\prime },b_{3}}^{{}}(\vec{k},\sigma )\right.  \nonumber \\   
&&-\left. \delta_{\sigma }^{\sigma ^{\prime }}   
T_{b_{1},b_{3}}^{\sigma^{\prime }}   
u_{\beta ,b_{2}}^{\ast }(\vec{k},\sigma )   
u_{\beta ^{\prime},b_{4}}^{{}}(\vec{k},\sigma )\right] \; ,  \label{7.16}   
\end{eqnarray}   
with    
\begin{mathletters}   
\label{7.16h}   
\label{Tmatrices}   
\begin{eqnarray}   
T_{b,b^{\prime }}^{\sigma } &\equiv &   
\frac{1}{L}\sum_{\vec{p},\beta }n_{\vec{p}}^{\beta \sigma }   
u_{\beta ,b}^{\ast }(\vec{p},\sigma )u_{\beta ,b^{\prime}}^{{}}(\vec{p},\sigma )   
\;,  \label{7.16b} \\   
n_{\vec{p}}^{\beta \sigma } &=&   
\left\langle \Psi _{0}^{\text{HF}}\left|    
\hat{a}_{\vec{p};\beta ,\sigma }^{+}   
\hat{a}_{\vec{p};\beta ,\sigma }^{{}}   
\right| \Psi _{0}^{\text{HF}}\right\rangle \;.   
\end{eqnarray}\end{mathletters}%   
Our Hartree-Fock theory is restricted to   
cases where the translational invariance  
is conserved but we still allow for a collinear ferromagnetic ground state.  
The Hartree-Fock Hamiltonian thus reads   
\begin{equation}   
\hat{H}^{\text{HF}} = \sum_{\vec{p};\beta,\sigma}    
E^{\text{HF}}_{\beta }(\vec{p},\sigma)   
\hat{a}_{\vec{p};\beta ,\sigma }^{+}   
\hat{a}_{\vec{p};\beta ,\sigma }^{{}} \; .   
\label{HFHamilt}   
\end{equation}   
In the special case where, (i), all orbitals~$b$ belong to different   
representations of the point-symmetry group of the lattice,   
and, (ii), the Hartree-Fock ground state    
$\left| \Psi_{0}^{\text{HF}}\right\rangle $ is invariant under   
the respective symmetry operations, Eq.~(\ref{7.16b}) becomes    
\begin{equation}   
T_{b,b^{\prime }}^{\sigma }=\delta_{b,b^{\prime }}n_{b,\sigma }^{0}\;.   
\label{7.16c}   
\end{equation}   
Under these conditions, the results   
of Ref.~\onlinecite{cook} are recovered; see below.   
   
\subsection{RPA from the equation-of-motion technique}   
   
\label{equa}   
   
We will derive the Green function (\ref{1.3}) in the random-phase   
approximation using the standard equation-of-motion technique\cite{Mahan}. 
First, we   
rewrite $G_{\text{ret}}(\vec{q},E)$ as    
\begin{mathletters}   
\label{7.5}   
\begin{eqnarray}   
G_{\text{ret}}(\vec{q},E) &=&   
\sum_{b,b^{\prime }} G_{\text{ret}}^{bb,b^{\prime }}(\vec{q},E)\;,   
\label{7.5a} \\   
G_{\text{ret}}^{b_{1}b_{2},b^{\prime }}(\vec{q},E)    
&\equiv &\frac{1}{L}\sum_{\vec{k},\vec{k}^{\prime }}   
\langle \langle    
\hat{c}_{\vec{k}+\vec{q};b_{1},\uparrow}^{+}   
\hat{c}_{\vec{k};b_{2},\downarrow }^{{}};   
\hat{c}_{\vec{k}^{\prime};b^{\prime },\downarrow }^{+}   
\hat{c}_{\vec{k}^{\prime }+\vec{q};b^{\prime},\uparrow }^{{}}   
\rangle \rangle_{E}\;.  \label{7.5b}   
\end{eqnarray}\end{mathletters}%   
The equation of motion for  
$G_{\text{ret}}^{b_{1}b_{2},b^{\prime }}(\vec{q},E)$   
cannot be decoupled self-consistently. Therefore, we will consider the more   
general Green function    
\begin{equation}   
\widetilde{G}_{\alpha_{1},\alpha_{2},b^{\prime}}^{\vec{k},\vec{k}^{\prime},\vec{q}}(E)   
\equiv \langle \langle    
\hat{a}_{\vec{k}+\vec{q};\alpha_{1},\uparrow }^{+}   
\hat{a}_{\vec{k};\alpha_{2},\downarrow }^{{}};   
\hat{c}_{\vec{k}^{\prime };b^{\prime },\downarrow }^{+}   
\hat{c}_{\vec{k}^{\prime }+\vec{q};b^{\prime },\uparrow }^{{}}   
\rangle \rangle _{E}\;,  \label{7.6}   
\end{equation}   
which allows us to express~(\ref{7.5b}) as    
\begin{equation}   
G_{\text{ret}}^{b_{1}b_{2},b^{\prime }}(\vec{q},E)=   
\frac{1}{L}\sum_{\alpha_{1},\alpha_{2}}   
\sum_{\vec{k}}   
u_{\alpha _{1},b_{1}}^{\ast }(\vec{k}+\vec{q},\uparrow )   
u_{\alpha _{2},b_{2}}^{{}}(\vec{k},\downarrow )   
\sum_{\vec{k}^{\prime }}   
\widetilde{G}_{\alpha_{1}\alpha_{2},b^{\prime}}^{\vec{k},\vec{k}^{\prime},\vec{q}}(E)\;.  \label{7.7}   
\end{equation}   
The equation of motion for    
$\widetilde{G}_{\alpha_{1}\alpha_{2},b^{\prime}}^{\vec{k},\vec{k}^{\prime },\vec{q}}(E)$ has the form    
\begin{mathletters}   
\label{7.8}   
\begin{eqnarray}   
E   
\widetilde{G}_{\alpha_{1}\alpha_{2},b^{\prime}}^{\vec{k},\vec{k}^{\prime},\vec{q}}(E) &=& A+B+C \; ,\\   
A &=& \langle [   
\hat{a}_{\vec{k}+\vec{q};\alpha_{1},\uparrow }^{+}   
\hat{a}_{\vec{k};\alpha _{2},\downarrow }^{{}}\,,\,   
\hat{c}_{\vec{k}^{\prime };b^{\prime },\downarrow }^{+}   
\hat{c}_{\vec{k}^{\prime }+\vec{q};b^{\prime },\uparrow }^{{}}  ]   
\rangle _{\Psi _{0}} \; , \label{7.8a} \\   
B &=&\langle \langle [   
\hat{a}_{\vec{k}+\vec{q};\alpha_{1},\uparrow }^{+}   
\hat{a}_{\vec{k};\alpha_{2},\downarrow }^{{}},   
\hat{H}_{1}]\,;\,   
\hat{c}_{\vec{k}^{\prime };b^{\prime },\downarrow }^{+}   
\hat{c}_{\vec{k}^{\prime }+\vec{q};b^{\prime },\uparrow }^{{}}   
\rangle \rangle_{E}  \; , 
\label{7.8b} \\   
C &=&\langle \langle [   
\hat{a}_{\vec{k}+\vec{q};\alpha_{1},\uparrow }^{+}   
\hat{a}_{\vec{k};\alpha _{2},\downarrow }^{{}},   
\hat{H}_{\text{at}}     ]\,;\,   
\hat{c}_{\vec{k}^{\prime };b^{\prime },\downarrow }^{+}   
\hat{c}_{\vec{k}^{\prime }+\vec{q};b^{\prime },\uparrow }^{{}}   
\rangle \rangle_{E}\;.  \label{7.8c}   
\end{eqnarray}\end{mathletters}%   
In RPA we assume that the exact ground state $\left| \Psi _{0}\right\rangle $   
of our Hamiltonian~(\ref{1.1}) may be replaced    
by the (spin-polarized) Hartree-Fock ground state    
$\left| \Psi _{0}^{\text{HF}}\right\rangle $ of~(\ref{HFHamilt}).   
Then, the expectation value~(\ref{7.8a}) becomes    
\begin{equation}   
A =\delta _{\vec{k},\vec{k}^{\prime }}u_{\alpha   
_{2},b^{\prime }}^{\ast }(\vec{k},\downarrow )u_{\alpha _{1},b^{\prime   
}}^{{}}(\vec{k}+\vec{q},\uparrow )\left( n_{\vec{k}+\vec{q}}^{\alpha   
_{1}\uparrow }-n_{\vec{k}}^{\alpha _{2}\downarrow }\right) \;.  \label{7.9}   
\end{equation}   
The evaluation of~(\ref{7.8b}) leads to    
\begin{equation}   
B =\sum_{\beta }E_{\alpha _{2},\beta }(\vec{k},\downarrow )%   
\widetilde{G}_{\alpha _{1},\beta ;b^{\prime }}^{\vec{k},\vec{k}^{\prime },\vec{q}%   
}(E)-\sum_{\beta }E_{\beta ,\alpha _{1}}(\vec{k}+\vec{q},\uparrow )\widetilde{G}%   
_{\beta ,\alpha _{2};b^{\prime }}^{\vec{k},\vec{k}^{\prime },\vec{q}}(E)\;.   
\label{7.10}   
\end{equation}   
Finally, the commutator in~(\ref{7.8c}) generates two-particle operators,   
which we decouple according to the rule   
\begin{equation}   
\langle \langle    
\hat{a}_{1\uparrow }^{+}   
\hat{a}_{2\uparrow }^{+}   
\hat{a}_{3\uparrow }^{{}}   
\hat{a}_{4\downarrow }^{{}};   
\,\ldots\rangle \rangle_{E}   
\approx    
\langle    
\hat{a}_{2\uparrow }^{+} \hat{a}_{3\uparrow }^{{}}   
\rangle_{\Psi _{0}^{\text{HF}}}   
\langle \langle \hat{a}_{1\uparrow }^{+}\hat{a}_{4\downarrow }^{{}};   
\,\ldots\rangle \rangle_{E}   
-\langle    
\hat{a}_{1\uparrow}^{+}\hat{a}_{3\uparrow }^{{}}   
\rangle_{\Psi_{0}^{\text{HF}}}   
\langle\langle \hat{a}_{2\uparrow }^{+}\hat{a}_{4\downarrow }^{{}};   
\,\ldots\rangle\rangle_{E}\;.  \label{7.11}   
\end{equation}   
This approximation appears to be natural because we replaced    
$\left| \Psi_{0}\right\rangle $    
by the one-particle product state $\left|\Psi_{0}^{\text{HF}}\right\rangle $.    
Note, however, the replacement~(\ref{7.11}) does not become   
exact even if we work with $\left|\Psi_{0}^{\text{HF}}\right\rangle $    
instead of $\left| \Psi_{0}\right\rangle $.   
   
After the application of the decoupling scheme~(\ref{7.11}),    
Eq.~(\ref{7.8c}) may be written as    
\begin{mathletters}   
\label{7.12}   
\begin{eqnarray}   
C &=& C_1 + C_2 + C_3 \\   
C_1 & = &    
\frac{2}{L}\sum_{\beta ,\beta ^{\prime }}\sum_{\vec{p};\sigma}   
\Bigl[n_{\vec{p}}^{\beta \sigma }   
V_{\beta\beta\sigma}^{\alpha_{2}\beta^{\prime}\downarrow}   
(\vec{p},\vec{k},\vec{k},\vec{p}\,)   
-n_{\vec{p}}^{\beta \downarrow }   
V_{\beta \beta ^{\prime}\downarrow }^{\alpha _{2}\beta \downarrow }   
(\vec{p},\vec{k},\vec{p},\vec{k}\,)\Bigr]   
\widetilde{G}_{\alpha_{1},\beta^{\prime};b^{\prime}}^{\vec{k},\vec{k}^{\prime},\vec{q}}(E) \; ,\label{7.12a} \\   
C_2 & = &    
- \frac{2}{L}\sum_{\beta ,\beta ^{\prime }}\sum_{\vec{p};\sigma }   
\Bigl[n_{\vec{p}}^{\beta \sigma }   
V_{\beta ^{\prime }\alpha_{1}\uparrow }^{\beta \beta \sigma }   
(\vec{k}+\vec{q},\vec{p},\vec{p},\vec{k}+\vec{q}\,)   
-n_{\vec{p}}^{\beta \uparrow }   
V_{\beta \alpha _{1}\uparrow}^{\beta ^{\prime }\beta \uparrow }   
(\vec{p},\vec{k}+\vec{q},\vec{p},\vec{k}+\vec{q}\,)\Bigr]    
\widetilde{G}_{\beta^{\prime},\alpha_{2};b^{\prime }}^{\vec{k},\vec{k}^{\prime },\vec{q}}(E)  \; , \nonumber \\   
\label{7.12b} \\   
C_3 & = &    
\frac{2}{L}\sum_{\beta ,\beta ^{\prime }}\sum_{\vec{p}}   
\left(    
n_{\vec{k}}^{\alpha _{2}\downarrow}   
-n_{\vec{k}+\vec{q}}^{\alpha _{1}\uparrow }\right)    
V_{\beta \alpha _{1}\uparrow }^{\alpha_{2}\beta ^{\prime }\downarrow }   
(\vec{p}+\vec{q},\vec{k},\vec{p},\vec{k}+\vec{q}\,)   
\widetilde{G}_{\beta,\beta^{\prime};b^{\prime}}^{\vec{p},\vec{k}^{\prime},\vec{q}}(E) \;.  \label{7.12d}   
\end{eqnarray}\end{mathletters}%   
When we use the explicit expression~(\ref{7.4}), the sum over~$\vec{p}$ can   
be carried out for $C_1$ and~$C_2$ because the Green functions    
do not depend on $\vec{p}$.    
\begin{mathletters}   
\label{7.14}   
\begin{eqnarray}   
C_1&=& 2\sum_{b_{1},b_{2},b_{3},b_{4},\sigma ,\sigma ^{\prime}}   
U_{\sigma \sigma^{\prime }}^{b_{1},b_{2},b_{3},b_{4}}   
\sum_{\beta}   
\delta_{\sigma ^{\prime }}^{\downarrow }   
\Bigl[    
\left(    
\delta _{\sigma}^{\downarrow }T_{b_{1},b_{4}}^{\downarrow }   
+T_{b_{1},b_{4}}^{\uparrow}\delta_{\sigma }^{\uparrow}   
\right)    
u_{\alpha _{2},b_{2}}^{\ast }(\vec{k},\downarrow )   
u_{\beta ,b_{3}}^{{}}(\vec{k},\downarrow )   
\nonumber \\   
&&     
\hphantom{2\sum_{b_{1},b_{2},b_{3},b_{4},\sigma ,\sigma ^{\prime}}   
U_{\sigma \sigma^{\prime }}^{b_{1},b_{2},b_{3},b_{4}}   
\sum_{\beta}   
\delta_{\sigma ^{\prime }}^{\downarrow }   
\Bigl[    
}    
-\delta_{\sigma }^{\downarrow }T_{b_{1},b_{3}}^{\uparrow}   
u_{\alpha _{2},b_{2}}^{\ast }(\vec{k},\downarrow )   
u_{\beta ,b_{4}}^{{}}(\vec{k},\downarrow )   
\Bigr]    
\widetilde{G}_{\alpha_{1},\beta;b^{\prime}}^{\vec{k},\vec{k}^{\prime},\vec{q}}(E)   
\label{7.14a} \; ,\\   
C_2&=& 2\sum_{b_{1},b_{2},b_{3},b_{4},\sigma ,\sigma ^{\prime}}   
U_{\sigma \sigma^{\prime }}^{b_{1},b_{2},b_{3},b_{4}}   
\sum_{\beta }\delta _{\sigma }^{\uparrow }   
\widetilde{G}_{\beta,\alpha_{2};b^{\prime}}^{\vec{k},\vec{k}^{\prime},\vec{q}}(E)   
\Bigl[    
T_{b_{1},b_{3}}^{\uparrow }\delta _{\sigma ^{\prime }}^{\uparrow}   
u_{\beta ,b_{2}}^{\ast }(\vec{k}+\vec{q},\uparrow )   
u_{\alpha_{1},b_{4}}^{{}}(\vec{k}+\vec{q},\uparrow )\nonumber \\   
&& \hphantom{2\sum_{b_{1},b_{2},b_{3},b_{4},\sigma ,\sigma ^{\prime}}   
U_{\sigma \sigma^{\prime }}^{b_{1},b_{2},b_{3},b_{4}}   
\sum_{\beta }\delta _{\sigma }^{\uparrow }   
\Bigl[ }   
-\left(   
\delta_{\sigma^{\prime}}^{\uparrow}T_{b_{2},b_{3}}^{\uparrow }   
+\delta_{\sigma^{\prime}}^{\downarrow}T_{b_{2},b_{3}}^{\downarrow }   
\right)    
u_{\beta ,b_{1}}^{\ast }(\vec{k}+\vec{q},\uparrow )   
u_{\alpha _{1},b_{4}}^{{}}(\vec{k}+\vec{q},\uparrow )   
\Bigr] \; , \nonumber \\   
\label{7.14b}\\   
C_3&=& - 2\sum_{b_{1},b_{2},b_{3},b_{4}}   
U_{\uparrow\downarrow}^{b_{1},b_{2},b_{3},b_{4}}   
\left(    
n_{\vec{k}+\vec{q}}^{\alpha _{1}\uparrow }   
-n_{\vec{k}}^{\alpha_{2}\downarrow }   
\right)    
u_{\alpha _{2},b_{2}}^{\ast }(\vec{k},\downarrow)   
u_{\alpha _{1},b_{4}}^{{}}(\vec{k}+\vec{q},\uparrow )\nonumber \\   
&& \hphantom{ - 2\sum_{b_{1},b_{2},b_{3},b_{4}}   
}   
\times \frac{1}{L}   
\sum_{\vec{p};\beta ,\beta ^{\prime }}   
u_{\beta ,b_{1}}^{\ast }(\vec{p}+\vec{q},\uparrow)   
u_{\beta ^{\prime },b_{3}}^{{}}(\vec{p},\downarrow )   
\widetilde{G}_{\beta ,\beta^{\prime };b^{\prime }}^{\vec{p},\vec{k}^{\prime },   
\vec{q}}(E)\; .   
\label{7.14c}   
\end{eqnarray}\end{mathletters}%   
With the help of the Hartree-Fock equations~(\ref{7.16})    
and relation~(\ref{7.7})    
we may cast the Green function~(\ref{7.6}) into the form    
\begin{eqnarray}   
\widetilde{G}_{\alpha_{1}\alpha_{2},b}^{\vec{k},\vec{k}^{\prime},\vec{q}}(E)   
&=&   
\left[ E-\left( E^{\text{HF}}_{\alpha _{2}}(\vec{k},\downarrow )-  
E^{\text{HF}}_{\alpha _{1}}(\vec{k}+\vec{q},\uparrow )\right) \right]^{-1}   
\nonumber \\   
&&   
\Bigl\{\delta_{\vec{k},\vec{k}^{\prime }}   
u_{\alpha _{2},b^{\prime }}^{\ast }(\vec{k},\downarrow )   
u_{\alpha _{1},b^{\prime }}^{{}}(\vec{k}+\vec{q},\uparrow)   
\left(    
n_{\vec{k}+\vec{q}}^{\alpha _{1}\uparrow }-n_{\vec{k}}^{\alpha_{2}\downarrow }   
\right)  \label{7.17}\\   
&&\hphantom{\Bigl\{   
}   
-2\!\!\sum_{b_{1},b_{2},b_{3},b_{4}}   
G_{\text{ret}}^{b_{1}b_{3},b^{\prime}}(\vec{q},E)   
U_{\downarrow \uparrow }^{b_{1},b_{2},b_{3},b_{4}}   
\left(    
n_{\vec{k}+\vec{q}}^{\alpha _{1}\uparrow }  
-n_{\vec{k}}^{\alpha _{2}\downarrow }   
\right)   
u_{\alpha _{2},b_{2}}^{\ast }(\vec{k},\downarrow )   
u_{\alpha _{1},b_{4}}^{{}}(\vec{k}+\vec{q},\uparrow )\Bigl\}\;.  \nonumber   
\end{eqnarray}   
This equation, together with~(\ref{7.7}), leads to    
\begin{equation}   
G_{\text{ret}}^{b_{1}b_{2},b^{\prime }}(\vec{q},E)=   
\left[\Gamma_{\text{ret}}(\vec{q},E)_{b^{\prime }b^{\prime}}^{b_{1}b_{2}}\right]  
+\sum_{b_{3},b_{4},b_{5},b_{6}}   
\left[\Gamma_{\text{ret}}(\vec{q},E)_{b_{3}b_{4}}^{b_{1}b_{2}}\right]  
\widetilde{U}_{b_{5}b_{6}}^{b_{3}b_{4}}   
G_{\text{ret}}^{b_{5}b_{6},b^{\prime }}(\vec{q},E)\;,   
\label{7.18}   
\end{equation}   
where    
\begin{mathletters}   
\label{7.19}   
\begin{eqnarray}   
\Gamma_{\text{ret}}(\vec{q},E)_{b_{3}b_{4}}^{b_{1}b_{2}}  
&\equiv &\frac{1}{L}   
\sum_{\vec{k},\alpha _{1},\alpha _{2}}   
\frac{u_{\alpha _{1},b_{1}}^{\ast }(\vec{k}+\vec{q},\uparrow )   
u_{\alpha _{2},b_{2}}^{{}}(\vec{k},\downarrow )   
u_{\alpha_{1},b_{3}}^{{}}(\vec{k}+\vec{q},\uparrow )   
u_{\alpha _{2},b_{4}}^{\ast }(\vec{k},\downarrow )   
}{   
E- \left(    
E^{\text{HF}}_{\alpha _{2}}(\vec{k},\downarrow )-   
E^{\text{HF}}_{\alpha_{1}}(\vec{k}+\vec{q},\uparrow )   
\right) +i\eta}   
\bigl(   
n_{\vec{k}+\vec{q}}^{\alpha _{1}\uparrow }-n_{\vec{k}}^{\alpha_{2}\downarrow }   
\bigr) \, , \nonumber\\   
&& \label{7.19a} \\   
\widetilde{U}_{b_{5}b_{6}}^{b_{3}b_{4}}    
&\equiv &   
-2U_{\downarrow \uparrow}^{b_{5},b_{4},b_{6},b_{3}}\;.  \label{7.19b}   
\end{eqnarray}\end{mathletters}%   
Here, we added the infinitesimal increment $\eta=0^+$ to ensure   
the properties of a retarded Green function~\cite{Mahan}.   
When we consider $\Gamma_{\text{ret}} (\vec{q},E)$    
and $\widetilde{U}$ in~(\ref{7.18}) as matrices with respect to the   
indices ${\frak {D}}=(b_{1}b_{2})$, the solution of Eq.~(\ref{7.7}) is given by    
\begin{equation}   
G_{\text{ret}}^{{\frak {D}},b^{\prime }}(\vec{q},E)   
=\sum_{{\frak {D}}^{\prime }}   
\left[   
\left(    
1-\Gamma_{\text{ret}} (\vec{q},E)\cdot    
\widetilde{U}   
\right)^{-1}   
\right]_{{\frak {D}}^{\prime }}^{{\frak {D}}}   
\Gamma_{\text{ret}}(\vec{q},E)_{b^{\prime }b^{\prime }}^{{\frak {D}}^{\prime }}  
\;.  \label{7.20}   
\end{equation}   
This result, together with Eq.~(\ref{7.5a}), gives us the Green function    
$G_{\text{ret}}(\vec{q},E)$, whose poles at $E^{\text{RPA}}(\vec{q}\,)$  
define the RPA spin-wave dispersion. We will further analyze  
it in section~\ref{comp}.    
   
\subsection{RPA from the diagrammatic approach}   
   
\label{diam}   
   
The result~(\ref{7.20}) of the RPA decoupling scheme may also be derived   
from the standard diagrammatic approach. Here, the matrix~(\ref{7.19a})    
is nothing but the retarded part of the respective {\em causal}   
transversal susceptibility $\Gamma$,    
\begin{equation}   
\Gamma_{b_{3}b_{4}}^{b_{1}b_{2}}(\vec{q},E)=   
\frac{1}{2\pi i}   
\sum_{\vec{k}}\int dE^{\prime }   
G_{b_{3}b_{1},\uparrow }^{\text{HF}}(\vec{q}+\vec{k},E+E^{\prime })   
G_{b_{2}b_{4},\downarrow }^{\text{HF}}(\vec{k},E^{\prime })     
\label{7.21} \; ,  
\end{equation}   
where the one-particle lines are evaluated within the Hartree-Fock   
approximation; see below.    
Using this expression and the vertex for spin-flip excitations   
$\widetilde{U}_{b_{5}b_{6}}^{b_{3}b_{4}}$ from~(\ref{7.19b}),   
we can calculate the causal two-particle Green function   
$G_{T}^{{\frak {D}},b^{\prime }}(\vec{q},E)$   
as the usual RPA sum of   
the ``bare bubbles'' in~(\ref{7.21}),   
\begin{mathletters}   
\label{7.22}   
\begin{eqnarray}   
G_{T}^{{\frak {D}},b^{\prime }}(\vec{q},E) &=&   
\Gamma_{b^{\prime}b^{\prime }}^{{\frak {D}}}(\vec{q},E)   
+   
\sum_{{\frak {D}}^{\prime },{\frak {D}}^{\prime\prime }}   
\Gamma_{{\frak {D}}^{\prime }}^{{\frak {D}}}(\vec{q},E)   
\widetilde{U}_{{\frak {D}}^{\prime \prime }}^{{\frak {D}}^{\prime }}   
\Gamma_{b^{\prime }b^{\prime }}^{{\frak {D}}^{\prime \prime }}(\vec{q},E)   
+\ldots \label{7.22a}   
\\   
&=&   
\Gamma_{b^{\prime }b^{\prime }}^{{\frak {D}}}(\vec{q},E)   
+\sum_{{\frak {D}}^{\prime },{\frak {D}}^{\prime \prime }}   
\Gamma_{{\frak {D}}^{\prime}}^{{\frak {D}}}(\vec{q},E)   
\widetilde{U}_{{\frak {D}}^{\prime \prime }}^{{\frak {D}}^{\prime }}   
G_{T}^{{\frak {D}}^{\prime \prime },b^{\prime }}(\vec{q},E) \; .   
\end{eqnarray}\end{mathletters}%   
This equation has the same solution as~(\ref{7.20})    
if we replace $\Gamma_{\text{ret}}$ and $G_{\text{ret}}$    
by $\Gamma$ and ${G}_{T}$, respectively. The   
retarded Green function $G_{\text{ret}}$ can then be re-derived    
from $G_{T}$ with the help of the standard relations\cite{Mahan}   
\begin{mathletters}   
\label{generalGF}   
\begin{eqnarray}   
\mathop{\rm Im}G_{\text{ret}}(\vec{q},E) &=&   
\mathop{\rm Im}G_{T}(\vec{q},E)    
\left(\Theta (E)-\Theta (-E)\right), \\   
\mathop{\rm Re}G_{\text{ret}}(\vec{q},E) &=&   
\mathop{\rm Re}G_{T}(\vec{q},E) \;.   
\end{eqnarray}\end{mathletters}%   
In order to show the equivalence of both approaches,   
it remains to calculate $\Gamma$ in~(\ref{7.21}) explicitly.    
To this end we first determine the   
Hartree-Fock Green function   
$G_{bb^{\prime },\sigma }^{\text{HF}}(\vec{k},E)$.    
Its diagrammatic evaluation leads to    
\begin{equation}   
G_{bb^{\prime },\sigma }^{\text{HF}}(\vec{k},E)=   
G_{bb^{\prime },\sigma }^{0}(\vec{k},E)+   
\sum_{b^{\prime }b^{\prime \prime }}   
G_{bb^{\prime },\sigma }^{0}(\vec{k},E)   
\Sigma_{b^{\prime }b^{\prime\prime },\sigma }^{\text{HF}}   
G_{bb^{\prime },\sigma }^{\text{HF}}(\vec{k},E)\;,  \label{7.23}   
\end{equation}   
where $G_{bb^{\prime},\sigma}^{0}(\vec{k},E)$   
is the causal Green function corresponding to   
the one-particle operator~$\hat{H}_1$ in~(\ref{1.1}),   
\begin{equation}   
\left[G_{\sigma}^{0}(\vec{k},E)\right]_{bb^{\prime}}^{-1}   
= E\delta_{b,b^{\prime}}-   
\varepsilon_{b,b^{\prime}}(\vec{k}) \; .   
\end{equation}   
Moreover, we introduced the usual Hartree-Fock self-energy    
\begin{equation}   
\Sigma_{bb^{\prime },\sigma }^{\text{HF}}=   
2\sum_{b_1 b_2,\sigma^{\prime }}   
\left(    
U_{\sigma ,\sigma ^{\prime }}^{b,b_1,b_2,b^{\prime }}   
T_{b_1,b_2}^{\sigma^{\prime }}   
-\delta_{\sigma }^{\sigma ^{\prime }}   
U_{\sigma ,\sigma }^{b,b_1,b^{\prime },b_2}   
T_{b_1,b_2}^{\sigma } \right)  \label{7.24}   
\end{equation}   
whose elements are calculated self-consistently,   
\begin{equation}   
T_{b,b^{\prime }}^{\sigma }=\sum_{\vec{k}}   
\left\langle \Psi_{0}^{\text{HF}}\left|    
\hat{c}_{\vec{k};(b\sigma )}^{+}   
\hat{c}_{\vec{k};(b^{\prime}\sigma )}^{{}}   
\right| \Psi_{0}^{\text{HF}}\right\rangle \; ,   
\end{equation}   
compare~(\ref{Tmatrices}).   
Using a matrix notation in~(\ref{7.23}) we find    
\begin{equation}   
\left[ G_{\sigma }^{\text{HF}}(\vec{k},E)\right]^{-1}   
=\left[ G_{\sigma }^{0}(\vec{k},E)\right] ^{-1}   
-\Sigma_{\sigma }^{\text{HF}}=   
E-\widetilde{E}_{\sigma }(\vec{k}) \label{7.25} \; ,   
\end{equation}   
where the matrix $\widetilde{E}_{\sigma }(\vec{k})$ is given by    
\begin{equation}   
\left( \widetilde{E}_{\sigma }(\vec{k})\right) _{b,b^{\prime }}=   
\varepsilon_{b,b^{\prime }}(\vec{k})+\Sigma_{bb^{\prime },\sigma }^{\text{HF}}\;.   
\label{7.26}   
\end{equation}   
{}From Eqs.~(\ref{7.25}) and~(\ref{7.26}) we see that    
$[G_{\sigma}^{\text{HF}}(\vec{k},E)]^{-1} $ is diagonalized    
by the matrix $u(\vec{k},\sigma )$ which   
solves the Hartree-Fock equations~(\ref{7.16}).   
The eigenvalues are given by $E^{\text{HF}}_{\beta }(\vec{k},\sigma )$.    
The diagonalization and inversion of~(\ref{7.25}) give the final result    
\begin{equation}   
\left( G^{\text{HF}}_{\sigma }(\vec{k},E)\right)_{\beta \beta ^{\prime}}=   
\delta_{\beta ,\beta^{\prime }}   
\frac{1}{E-E^{\text{HF}}_{\beta}(\vec{k},\sigma )+i\eta_{\beta }(\vec{k},\sigma )}   
\;.  \label{7.27}   
\end{equation}   
Again, we added a positive or negative increment $\eta_{\beta }(\vec{k},\sigma )$    
depending on whether $E^{\text{HF}}_{\beta }(\vec{k},\sigma )$ is   
larger or smaller than the Fermi energy. This ensures the analytical   
properties of a causal Green function. We insert~(\ref{7.27})   
into~(\ref{7.21}), and find   
\begin{eqnarray}   
\Gamma_{b_{3}b_{4}}^{b_{1}b_{2}}(\vec{q},E) &=&   
\frac{1}{L}\sum_{\vec{k},\alpha_{1},\alpha _{2}}   
u_{\alpha _{1},b_{1}}^{\ast }(\vec{k}+\vec{q},\uparrow )   
u_{\alpha _{2},b_{2}}^{{}}(\vec{k},\downarrow )   
u_{\alpha_{1},b_{3}}^{{}}(\vec{k}+\vec{q},\uparrow )   
u_{\alpha _{2},b_{4}}^{\ast }(\vec{k},\downarrow )\nonumber \\   
&&\times \biggl(    
\frac{(1-n_{\vec{k}+\vec{q}}^{\alpha _{1}\uparrow })   
n_{\vec{k}}^{\alpha _{2}\downarrow }   
}{   
E-\left( E^{\text{HF}}_{\alpha _{1}}(\vec{k}+\vec{q},\uparrow)    
-E^{\text{HF}}_{\alpha _{2}}(\vec{k},\downarrow )\right) +i\eta}     
\\   
&& \hphantom{\biggl(   
}   
-    
\frac{(1-n_{\vec{k}}^{\alpha _{2}\downarrow })   
n_{\vec{k}+\vec{q}}^{\alpha_{1}\uparrow }   
}{   
E-\left( E^{\text{HF}}_{\alpha _{1}}(\vec{k}+\vec{q},\uparrow    
-E^{\text{HF}}_{\alpha _{2}}(\vec{k},\downarrow ))   
\right) -i\eta}\biggr) \;. \nonumber   
\end{eqnarray}   
{}From this equation one easily re-derives~(\ref{7.19a}) with the help   
of the general relations~(\ref{generalGF}). In this way, we have shown the   
complete equivalence of the diagrammatic and the equation-of-motion    
derivation of the multi-band RPA equations.   
   
\section{Spin-wave dispersions}   
\label{comp}   
   
\subsection{Variational spin-wave dispersion}   
   
In Ref.~\onlinecite{PRB98} we proposed the following Gutzwiller wave-function\cite{gutzi} for a   
variational examination of the Hamiltonian (\ref{1.1}):    
\begin{equation}   
|\Psi _{\text{G}}\rangle =\hat{P}_{\text{G}}|\Phi _{0}\rangle \;.   
\label{GWF}   
\end{equation}   
Here, $|\Phi _{0}\rangle $ is any normalized single-particle product-state   
and the Gutzwiller correlator $\hat{P}_{\text{G}}$ allows for a variational   
adjustment of the occupation of local atomic multiplets. Our choice for the   
correlator $\hat{P}_{\text{G}}$ ensures that the wave-function (\ref{GWF})   
yields the exact ground state of $\hat{H}$ both in the uncorrelated and the   
atomic limit. For all other values of correlation parameters the expectation   
value of $\hat{H}$ in the wave-function can be determined analytically in   
the limit of large spatial dimensions, $d\to\infty$.  
As in our recent work, this limit must   
be considered as yet another approximation since  
we apply our general analytical   
results to real three-dimensional systems.  Note that $1/d$-corrections 
are expected to be small~\cite{gutzjubel,GebPRB90}. 
   
Once the optimum variational  
ground state $|\Psi_{\text{G}}^{\text{opt}}\rangle$    
has been found by minimizing the ground-state energy-functional,   
the following expression for the variational spin-wave dispersion can be   
evaluated    
\begin{equation}   
E^{\text{var}}(\vec{q}\,) =  
\frac{   
\left\langle \Psi _{\text{G}}^{\text{opt}}\left|    
\hat{S}_{\vec{q}}^{+}\hat{H}\hat{S}_{\vec{q}}^{-}   
\right| \Psi_{\text{G}}^{\text{opt}}\right\rangle    
}{   
\left\langle \Psi_{\text{G}}^{\text{opt}}\left|    
\hat{S}_{\vec{q}}^{+}\hat{S}_{\vec{q}}^{-}   
\right|\Psi _{\text{G}}^{\text{opt}} \right\rangle}-   
\frac{   
\langle\Psi _{\text{G}}^{\text{opt}} | \hat{H}| \Psi _{\text{G}}^{\text{opt}}   
\rangle    
}{   
\langle\Psi _{\text{G}}^{\text{opt}} |\Psi _{\text{G}}^{\text{opt}}    
\rangle }\;.    
\end{equation}   
In Ref.~\onlinecite{spinpaper} a general    
analytical expression for $E^{\text{var}}(\vec{q}\,)$    
has been found in the limit of large spatial dimensions.    
The numerical evaluation of this   
result for real materials like iron or nickel is involved but feasible.   
So far, we have applied our method only to a degenerate two-band model. The   
results for this model will be compared with the RPA in the next subsection.   
   
\subsection{Two-band model}   
   
In Ref.~\onlinecite{spinpaper} we have   
discussed the variational spin-wave dispersion for   
a model with two degenerate $e_{g}$ orbitals ($b=1,2$)   
on a simple-cubic lattice.    
In this system the general   
atomic Hamiltonian~(\ref{1.2}) becomes    
\begin{eqnarray}   
\hat{H}_{\text{at}} &=&U\sum_{b}\hat{n}_{b,\uparrow }\hat{n}_{b,\downarrow   
}+U^{\prime }\sum_{\sigma ,\sigma ^{\prime }}\hat{n}_{1,\sigma }\hat{n}%   
_{2,\sigma ^{\prime }}-J\sum_{\sigma }\hat{n}_{1,\sigma }\hat{n}_{2,\sigma }   
\\[3pt]   
&&+J\sum_{\sigma }\hat{c}_{1,\sigma }^{+}\hat{c}_{2,-\sigma }^{+}\hat{c}%   
_{1,-\sigma }^{\vphantom{+}}\hat{c}_{2,\sigma }^{\vphantom{+}}+J_{\text{C}%   
}\Bigl(\hat{c}_{1,\uparrow }^{+}\hat{c}_{1,\downarrow }^{+}\hat{c}%   
_{2,\downarrow }^{\vphantom{+}}\hat{c}_{2,\uparrow }^{\vphantom{+}}+\hat{c}%   
_{2,\uparrow }^{+}\hat{c}_{2,\downarrow }^{+}\hat{c}_{1,\downarrow }^{%   
\vphantom{+}}\hat{c}_{1,\uparrow }^{\vphantom{+}}\Bigr)\;.  \nonumber   
\end{eqnarray}   
In cubic symmetry, the Coulomb and exchange-integrals $U$, $U^{\prime }$,    
$J$ and $J_{\text{C}}$ are not independent from each other. Instead we have   
only two free parameters, because the relations $J=J_{\text{C}}$ and    
$U-U^{\prime }=2J$ hold.   
   
In Ref.~\onlinecite{PRB98} we have    
discussed the appearance of ferromagnetic order in   
this model in detail both for the Hartree-Fock and our Gutzwiller theory. As   
is well known for mean-field theories, the stability of a ferromagnetic   
solution is drastically overestimated in a Hartree-Fock treatment.    
In a Hartree-Fock description only the Stoner parameter $I=(U+J)/2$   
governs the magnetic behavior. The paramagnet becomes unstable if the   
Stoner-criterion $I\,N_{E_{\text{F}}}>1$ is fulfilled,   
where $N_{E_{\text{F}}}$    
is the density of states at the Fermi level. This means that a   
ferromagnetic transition occurs for any value of the local exchange   
constant~$J$.   
This is in striking contrast to our variational approach.   
In the Gutzwiller many-body approach it becomes evident   
that a sizeable Hund's-rule coupling~$J$ is crucial    
for the formation of ferromagnetic order.   
Thus, for small values of $J$, the Hartree-Fock theory leads to   
qualitatively incorrect results.    
   
The significant differences for the critical values of the Coulomb interaction    
complicate a comparison between the RPA and the   
variational spin-wave approach. For the same set of parameters $J$ and $U$   
the underlying ground states are completely different and, therefore, a   
comparison of the spin-wave properties would not make sense. Thus, it   
appears to be more reasonable to consider the results of both methods for   
the same magnetization per band,   
\begin{equation}   
m= \frac{1}{L} \sum_l \langle \hat{S}^z_{l;b}\rangle =   
\frac{1}{2L} \sum_l \langle \hat{n}_{l;b\uparrow}\rangle    
- \langle \hat{n}_{l;b\downarrow}\rangle  \; .   
\end{equation}   
The density of electrons per band and spin direction  
in the paramagnetic phase is given by   
$n=(1/L) \sum_{l} \langle \hat{n}_{l;b,\sigma}\rangle $.   
   
\begin{figure}[tbp]    
\epsfig{figure=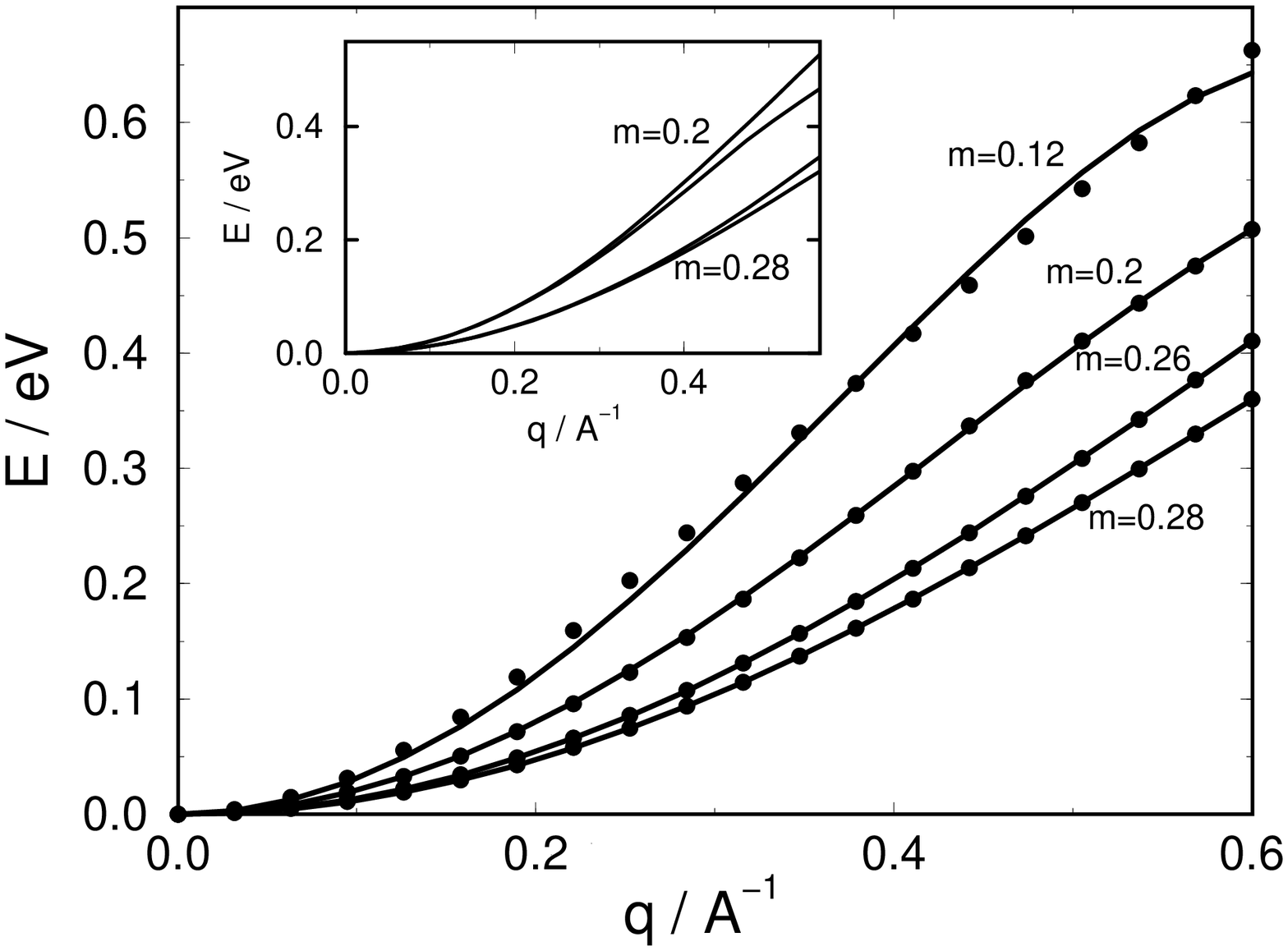,width=10cm}    
\caption{Variational spin-wave dispersion in $x$~direction,  
$E^{\text{var}}((q,0,0))$,  
for a generic two-band model with $n=0.3$, $J=0.2U$, and the values   
$U/{\rm eV}=7.8,\, 10,\, 12,\, 13.6$   
correspond to $m=0.12,\, 0.20,\, 0.26,\, 0.28$.  
The lattice constant is $a=2.5\,$\AA.\\   
Inset:   
$E^{\text{var}}((q,0,0))$ and   
$E^{\text{var}}((q/\sqrt{2},q/\sqrt{2},0))$  
for $m=0.2$ and $m=0.28$, respectively. The spin-wave dispersion is   
almost isotropic for strong ferromagnets.}  
\label{fig1}    
\end{figure}

In Fig.~\ref{fig1} the variational spin-wave dispersion in $x$-direction   
is shown for four different magnetizations. Here we used the same   
tight-binding parameters as in our analysis of the ground-state properties   
in~\onlinecite{PRB98}. The average electron density per orbital   
and spin direction in the paramagnet is $n=0.3$.    
We keep the ratio $J/U=0.2$ fixed and consider different   
values $U/{\rm eV}=7.8,$ $10,$ $12,$ $13.6$ (bandwidth $W=6.6~{\text eV}$)   
which correspond to a   
magnetization per band of $m=0.12,$ $0.20,$ $0.26,$ $0.28$. The last value   
belongs to the almost fully polarized ferromagnet ($m\approx n$).   
As can be seen from Fig.~\ref{fig1}, the   
spin-wave dispersion drastically depends on the magnetization. The data can   
be fitted very well to   
\begin{equation}   
E^{\text{var}}((q_{x},0,0))= Dq_{x}^{2}(1+\beta q_{x}^{2})   
+{\cal O}(q_{x}^{6})\;,   
\end{equation}   
in qualitative agreement with experiments on iron-group metals. In the case   
of a strong ferromagnet, we obtain $D=1.4\,\text{eV\AA$^{2}$}$ and    
$D=1.2\,\text{eV\AA$^{2}$}$ for $m=0.26$ and $m=0.28$, respectively.    
This is the right   
order of magnitude in nickel where $D=0.43\,\text{eV\AA$^{2}$}$.    
The inset of Fig.~\ref{fig1} shows that the spin-wave dispersion is   
almost isotropic which also agrees with experimental observations.   
In our variational many-body approach, the spin-wave stiffness    
{\em decreases\/} as a function of~$U$. This does not come as a surprise   
because the effective coupling between the sites decreases   
when the electron hopping between the sites becomes less effective.   
   
\begin{figure}[tbp]    
\epsfig{figure=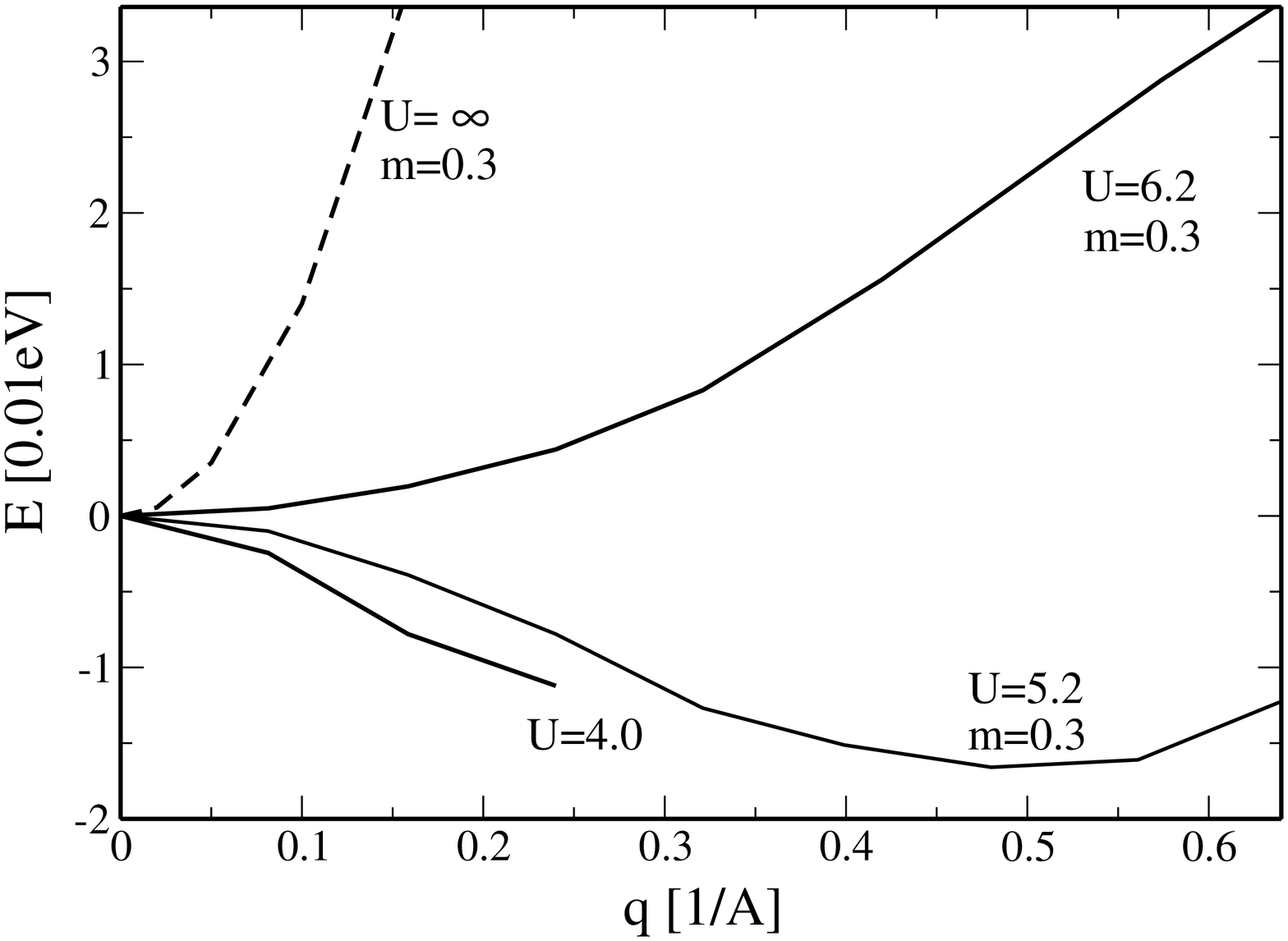,width=10cm,angle=-90}    
\caption{RPA spin-wave dispersion   
in $x$~direction, $E^{\text{RPA}}((q,0,0))$,  
for a generic two-band model with $n=0.3$ and $J=0.2U$. The value   
$U=4.0\, \text{eV}$ corresponds to $m<n$ in Hartree-Fock theory  
whereas $U=5.2\, \text{eV}$ and $U=6.2\, \text{eV}$ correspond to   
the fully polarized Hartree-Fock ground state.  
Also shown is the spin-wave dispersion for $U=\infty$.  
The lattice constant is $a=2.5\,$\AA.}    
\label{fig2}    
\end{figure}

In Fig.~\ref{fig2}    
the results for the spin-wave dispersion in the RPA are shown for   
three different values of~$U$. In contrast to our variational approach, the   
RPA predicts a {\em negative\/} slope of the dispersion    
for non-saturated ferromagnetism, $m<n$.  
Note that for $U=5.2\,{\rm eV}$ the system has just attained the fully   
polarized state (compare Fig.~7 of Ref.~\onlinecite{PRB98}).    
But even in this case,   
a well-defined RPA spin-wave excitation does not yet exist.   
It requires a further increase of~$U$   
until the fully polarized collinear Hartree-Fock  
ground state appears to be stable.  
Since the RPA probes the `local' stability of a  
mean-field state it is conceivable 
that another ferromagnetic Hartree-Fock state with some spiral order 
has a lower energy than the collinear state.  
However, ferromagnetic  
phases with spiral order are not generic for transition metals. 
We argue that this prediction of exotic ferromagnetic spin structures 
is an artifact of the Hartree-Fock approach.  
In fact, as we have demonstrated for our two-band model~\cite{PRB98}, 
Hartree-Fock theory predicts ferromagnetism 
in regions of the $(U,J)$~phase diagram where the paramagnetic phase 
is actually stable.  
 
In the case of a stable collinear ferromagnetic Hartree-Fock  
ground state, the RPA spin-wave stiffness   
{\em increases\/} as a function of~$U$, again in   
contrast to our variational approach.  
For all generic cases,   
$D^{\text{RPA}}(U\to\infty) \geq D^{\text{var}}(U\to\infty)$ holds   
in the limit of infinite coupling.   
In our case, $D^{\text{RPA}}(U\to\infty) =1.4\, \text{eV\AA$^2$}$,  
as compared to $D^{\text{var}}(U\to\infty)=0.4\, \text{eV\AA$^2$}$.  
The variational and RPA results differ significantly   
in the strong-coupling limit, too.  
Nevertheless, there are some intermediate  
coupling strengths $U_0$, $J_0$ where the variational   
and the RPA spin-wave stiffness agree. Such accidental  
agreements may also occur when RPA results are compared to   
experimental data.  
  
In some special cases where the fully   
polarized state is a one-particle product state   
we find $D^{\text{RPA}}(U,J) < D^{\text{var}}(U,J)$    
for all finite values of $U$, $J$, and $D^{\text{RPA}}(U\to\infty) \to   
D^{\text{var}}(U\to\infty)$ for large couplings.   
This is also the case in the one-band system; see below.   
   
\subsection{One-band model}   
   
The study of the one-band model allows us to   
elucidate the origin of the qualitatively different   
results for the spin-wave dispersion.   
In the case of the one-band model,  
\begin{equation}   
\hat{H}=\sum_{i,j}\sum_{\sigma ,\sigma^{\prime }}t_{i,j}\hat{c}_{i;\sigma}^{+}   
\hat{c}_{j;\sigma }^{\vphantom{+}}+U\sum_{i}\hat{n}_{i\uparrow }   
\hat{n}_{i\downarrow }=\sum_{\vec{k}}   
\varepsilon (\vec{k})\,c_{\vec{k};\sigma}^{+}c_{\vec{k};\sigma }^{\vphantom{+}}+ U\sum_{i}\hat{n}_{i\uparrow }   
\hat{n}_{i\downarrow } \; ,  
\end{equation}   
the results of Sect.~\ref{rand} simplify significantly. The Hartree-Fock   
energies can be written as    
\begin{equation}   
E^{\text{HF}}(\vec{k},\sigma )=\varepsilon (\vec{k})+Un_{-\sigma }   
\end{equation}   
and the susceptibility~(\ref{7.19a}) becomes    
\begin{mathletters}  
\begin{eqnarray}   
\Gamma_{\text{ret}} (\vec{q},E)    
&=&\frac{1}{L}\sum_{\vec{k}}\frac{n_{\vec{k}+\vec{q}\uparrow }   
-n_{\vec{k}\downarrow }}{E-\Delta -\left( \varepsilon (\vec{k})-   
\varepsilon (\vec{k}+\vec{q}\,)\right) +i\eta }\;,  \label{8.5} \\   
\Delta &=&U(n_{\uparrow }-n_{\downarrow })\;.   
\end{eqnarray}\end{mathletters}%  
For the spin-wave energies $E^{\text{RPA}}(\vec{q\,})$   
the denominator in    
\begin{equation}   
G_{\text{ret}}(\vec{q},E)=   
\frac{\Gamma_{\text{ret}} (\vec{q},E)}{1+U\,\Gamma_{\text{ret}} (\vec{q},E)}   
\end{equation}   
vanishes, i.e.,   
\begin{equation}   
1+U\,\Gamma_{\text{ret}} (\vec{q},E^{\text{RPA}}(\vec{q}\,))=0\;.   
\end{equation}   
To the leading (second) order in $\vec{q}$ we can expand~(\ref{8.5}) in   
powers of $1/\Delta $ because   
\begin{equation}   
|E-( \varepsilon (\vec{k})-\varepsilon (\vec{k}+\vec{q\,})) |\ll   
\Delta    
\end{equation}   
is fulfilled for our low-lying spin-wave excitations,   
see~(\ref{1.6}).   
This expansion leads to    
\begin{eqnarray}   
E^{\text{RPA}}(\vec{q}\,)&=&   
\frac{1}{(n_{\uparrow }-n_{\downarrow })L}   
\biggl[   
\sum_{\vec{k}}(n_{\vec{k}+\vec{q}\uparrow}-n_{\vec{k}\downarrow })   
\left(   
\varepsilon (\vec{k})-\varepsilon (\vec{k}+\vec{q\,})   
\right) \nonumber \\   
&& -\frac{1}{\Delta}   
\sum_{\vec{k}}(n_{\vec{k}+\vec{q}\uparrow }-n_{\vec{k}\downarrow })   
\left(\varepsilon (\vec{k})-\varepsilon (\vec{k}+\vec{q\,})\right)^{2}\biggr] \; .   
\label{8.8}   
\end{eqnarray}   
This result is to be compared with our variational spin-wave dispersion.    
However,   
such a comparison is somewhat questionable, since the existence of a stable   
ferromagnetic phase is hard to obtain in our variational approach.    
Nevertheless, let us assume that a given one-band model with   
a special density of states may have a fully-polarized   
ferromagnetic ground state. Then we obtain the following expression for the   
variational spin-wave dispersion    
\begin{equation}   
E^{\text{var}}(\vec{q}\,)=\frac{1}{n_{\uparrow }L}   
\sum_{\vec{k}}n_{\vec{k}+\vec{q}\uparrow }   
\left( \varepsilon (\vec{k})-\varepsilon (\vec{k}+\vec{q\,})\right) \;.   
\end{equation}   
This expression differs from $E^{\text{RPA}}(\vec{q}\,)$    
in~(\ref{8.8}) by the second term, i.e., for $n_{\downarrow }=0$ we find    
\begin{equation}   
\delta E =E^{\text{RPA}}(\vec{q}\,)-E^{\text{var}}(\vec{q}\,)=  
-\frac{1}{\Delta }\frac{1}{n_{\uparrow }L}   
\sum_{\vec{k}}n_{\vec{k}\uparrow }\left(   
\varepsilon (\vec{k})-\varepsilon (\vec{k}+\vec{q}\,)\right)^{2}\;.    
\end{equation}   
The energy difference $\delta E$ is always {\em negative\/} and vanishes for   
$U\to\infty$, i.e., for $\Delta \to\infty $. It is essentially this   
contribution which eventually leads to negative values of the RPA   
spin-wave stiffness for finite~$U$ in the two-band model.   
   
{}From a formal point of   
view the origin of the   
differences between both theories is obvious:   
For a fully polarized ground state of the one-band model the   
variational dispersion is identical to the {\em exact\/} first moment of the   
spectral function $\chi_{\text{ret}}(\vec{q},E)=   
\mathop{\rm Im}G_{\text{ret}}(\vec{q},E)$.   
However, the RPA predicts the existence of spectral   
weight around $E=\Delta $,    
the so-called ``Stoner-excitations''. Thus, the spin-wave pole of the RPA   
must have a lower energy than $E^{\text{var}}(\vec{q}\,)$, since, otherwise,   
the RPA could not give the correct first moment of $\chi_{T}(\vec{q},E)$.   
For this reason we need to address the issue of whether or not the   
RPA-theory correctly describes    
the {\em high-energy physics\/} of itinerant ferromagnets.   
   
First, it should be noted that the {\em existence\/} of the  
energy scale $\Delta$ is an artifact  
of the Hartree-Fock theory which survives in the RPA. From    
ground-state concepts like the magnetic condensation energy we infer   
that such an energy scale is an artificial feature of mean-field theories,   
and irrelevant within a many particle description of itinerant    
ferromagnets~\cite{PRB98}.   
Second, the misleading {\em relevance\/} of this energy scale stems from   
the overestimation of longitudinal fluctuations in mean-field theories.    
   
In order to see the second point,   
we consider again the spectral function $\chi_{T}(\vec{q},E)$,~(\ref{swf}),   
which provides the energy distribution of the spin excitations.   
In a spin-wave state the local moments do not depend on momentum, i.e,    
$\langle \Sqi \rangle_{\Psi^{0}_{\vec{q}}}=\langle \Sqi    
\rangle_{\Psi_{0}}$. In particular, this implies for the fully polarized    
one-band model that there are no doubly occupied    
lattice sites in $|\Psi^{0}_{\vec{q}}\rangle$.    
The temporal development of this state is now crucial:   
In a strongly correlated many-particle system,   
longitudinal fluctuations are substantially suppressed because charge   
fluctuations are energetically too costly at large~$U$.   
Therefore, the length of local spins $\langle \Sqi \rangle$    
is basically conserved also as a function of time.    
Our variational method correctly captures this generic   
behavior:   
fast longitudinal fluctuations are small and may be neglected at least    
for small values of~$\vec{q}$. In contrast, longitudinal fluctuations   
are always present in mean-field theory   
since electrons with opposite spin meet each other with non-zero    
probability. Recall that Hartree-Fock theory predicts a ferromagnetic  
ground state for much smaller interaction strengths than  
a proper many-body treatment.  
The presence and importance of   
considerable longitudinal spin fluctuations within Hartree-Fock and RPA  
then leads to   
the prediction of an instability of the   
collinear ferromagnetic phase   
($D^{\text{RPA}}<0$), as seen in Fig.~\ref{fig2}.   
   
These findings also explain the qualitatively different behavior    
of the spin-wave stiffness as a function of~$U$.    
In the variational approach we find a stable ferromagnetic ground state    
with well-defined local moments. The interaction of these `spins'    
is mediated by hopping processes. Similar to the well-known large-$U$    
antiferromagnetic coupling $J_{\text{AF}}\propto t^2/U$ in the case of    
the half-filled Hubbard model, we find that the effective    
spin-coupling $J_{\text{F}}$ in ferromagnets is reduced with increasing~$U$   
because hopping processes are more and more suppressed as they induce   
charge fluctuations. Note that the coupling $J_{\text{F}}$    
remains finite in the limit $U\to\infty$ because we are dealing    
with a non-integer band-filling    
where electron transfers are never completely suppressed.   
In contrast to this generic behavior,   
the Hartree-Fock theory misleadingly predicts   
a ferromagnetic ground state already for rather small interaction strengths   
where local moments are only partially generated.    
The RPA dynamics of the `Hartree-Fock spins' is considerably determined by    
longitudinal fluctuations. With increasing interaction strength    
the local moments are stabilized even within Hartree-Fock theory.   
Consequently, the RPA spin-wave stiffness increases and eventually   
overshoots the value from our variational approach.   
The basic mechanism for a reduction of $J_{\text{F}}$   
in our variational description remains completely hidden in the RPA   
because the reduction of the electron transfer amplitudes  
due to the electrons' mutual Coulomb interaction   
is not contained in a Hartree-Fock mean-field approach.  
   
\section{Conclusions}   
\label{summ}   
   
In this work we have derived the random-phase approximation (RPA)   
for the analysis of spin-wave properties in general   
multi-band Hubbard models with a collinear ferromagnetic ground state.   
The equation-of-motion technique and the diagrammatic approach  
lead to identical expressions.  
  
We have applied our analytical results to model systems   
with one and two correlated orbitals per lattice site.   
This allows us to compare the RPA results with those of  
a recently introduced variational approach to spin-wave excitations  
of Gutzwiller-correlated many-body ground states.  
The numerical analysis of the two-band model   
reveals significant qualitative differences   
between both methods. In the RPA,   
the spin-wave stiffness is negative for non-saturated ferromagnetism.  
With increasing correlations the RPA spin-wave energies are shifted upwards  
and the RPA spin-wave stiffness eventually becomes positive in the saturated   
ferromagnet. Our variational correlated-electron approach   
shows a completely different behavior.   
The spin-wave energies are always positive and the spin-wave stiffness  
decreases with increasing interactions.   
The same discrepancies occur in the one-band case where a fully  
analytical evaluation of the spin-wave dispersions is possible.  
  
We conclude that the Hartree-Fock/RPA   
theory provides an inadequate description   
for spin-wave properties in correlated electron systems.   
In our view, the main problem is the mean-field character   
of the underlying Hartree-Fock theory. In this approach   
the ferromagnetic ground state lacks well-developed local moments,  
and, therefore, high-energy longitudinal fluctuations  
of the local spins are large and relevant.   
As a consequence, RPA theory predicts   
an instability of the collinear ferromagnetic ground state  
against some spiral order.  
Our many-body approach reveals quite a different picture.  
In the region of stable ferromagnetism we find well-defined local moments.   
Spin excitations in these systems are transversal fluctuations   
of the local moments whose length is essentially conserved in space and 
time.  
Consequently, the stiffness of the variational  
spin-wave dispersion is always positive and decreases for   
increasing interaction strength.  
  
The popular ``frozen-magnon'' approximation for the examination   
of spin-wave excitations in itinerant ferromagnets   
can be considered as a mean-field theory  
in which the conservation of the local spins is introduced by hand.    
In this way, the overestimation of longitudinal fluctuations in the 
Hartree-Fock/RPA approach is eliminated by construction.   
In fact, it can be shown that the frozen-magnon    
approximation would lead to the same results as our variational approach    
if we chose a Hartree-Fock variational wave-function    
instead of our Gutzwiller wave-function. 
A more detailed analysis of the frozen-magnon approach  
will be the subject of a forthcoming study~\cite{wbp}.

\acknowledgments  
  
We thank Werner Weber and David Logan for helpful discussions.

\newpage  
  
\begin{references}  
  
\bibitem{stoner} Stoner E C 1938 Proc.~Roy.~Soc.~A~{\bf 165} 372;    
 Slater J C 1953 Rev.\ Mod.\ Phys.~{\bf 25} 199;   
Wohlfarth E P 1953 {\sl ibid.}, 211    
  
\bibitem{vanVleck} Van Vleck J H 1953  Rev.\ Mod.\ Phys.~{\bf 25} 220  
  
\bibitem{gutzi} Gutzwiller M C 1963 Phys.~Rev.~Lett.~{\bf 10},   
159; 1964 Phys.~Rev.~{\bf 134}, A923;   
1965 Phys.~Rev.~{\bf 137} A1726   
  
\bibitem{Herring} Herring C 1966 Magnetism Vol.~IV, ed. by  
G T Rado and H Suhl (New York: Academic Press) p.~1  
  
\bibitem{moriya} Morija T 1985 Spin Fluctuations in Itinerant Electron  
Magnetism (Berlin: Springer)  
  
\bibitem{Capellmann} Capellmann H (Ed) 1986 Metallic Magnetism   
(Berlin: Springer)  
  
\bibitem{yosida} Yosida K 1998 Theory of Magnetism (Berlin: Springer)  
  
\bibitem{Fazekas} Fazekas P 1999 Lecture Notes on Electron Correlation  
and Magnetism (Singapore: World Scientific)  
  
\bibitem{Eberhardt} Eberhardt W and Plummer E W 1980 Phys. Rev. B {\bf 21}  
3245  
  
\bibitem{gutzjubel} B\"unemann J, Gebhard F and Weber W 2000   
Found. of Physics {\bf 30} 2011   
  
\bibitem{Vollhardtrev} For a recent review, see Vollhardt D,  
Bl\"umer N, Held K, Kollar M, Schlipf J, Ulmke M and  
Wahle J 1999 Adv. in Solid-State Physics {\bf 28} 383  
  
\bibitem{ulmke} Ulmke M 1998 Eur.~Phys.~J.~B~{\bf 1} 301;   
Tasaki H 1998 Prog.~Theor.~Phys.~{\bf 99} 489  
  
\bibitem{PRB98} B\"{u}nemann J, Weber W and   
Gebhard F 1998 Phys.\ Rev.\ B~{\bf 57}, 6896  
  
\bibitem{vollhardt} Held K and   
Vollhardt D 1998 Eur.~Phys.~J.~B~{\bf 5}, 473   
  
\bibitem{Reinhard} Daul S and Noack R M 1998 Phys.~Rev.~B~{\bf 58} 2635;  
Guerro M and Noack R M 2001 Phys.~Rev.~B~{\bf 63} 144423  
  
\bibitem{Japaner} Sakamoto H, Momoi T and Kubo K 2001 (unpublished;  
arXiv cond-mat/0104223)  
  
\bibitem{JPhys} B\"unemann J, Gebhard F and Weber W 1997 J. Phys. Cond.   
Matt. {\bf 8} 7343   
  
\bibitem{wandlitz} Nolting W (Ed) (forthcoming)  
Ground-State and Finite-Temperature  
Bandferromagnetism (Berlin: Springer)  
  
\bibitem{lovesey} Marshall W and Lovesey S W 1971 Theory of Thermal   
Neutron Scattering (Oxford: Oxford University Press)   
  
\bibitem{mook} Mook H A, Nicklow R M,   
Thompson E D and Wilkinson M K 1969   
J. Appl. Phys. {\bf 40} 1450;   
Lowde R D and Windsor C G 1970 Adv. Phys. {\bf 19} 813  
  
\bibitem{cook} Cooke J F, Lynn J W and Davis H L 1980   
Phys.~Rev.\ B~{\bf 21} 4118    
  
\bibitem{mills} Tang H, Plihal M and Mills D L 1998   
J. Magn. Magn. Matt.~{\bf 187} 23;   
Plihal M and D L Mills 1998   
Phys. Rev. B~{\bf 58} 14407;   
Hong J and Mills D L 2000 Phys. Rev. B~{\bf 61}, R858  
  
\bibitem{eschrig} Uhl M and K\"ubler J 1996   
Phys. Rev. Lett. {\bf 77} 334; Rosengaard N M and Johansson B 1997   
Phys. Rev. B {\bf 55} 14975; Halilov S V, Eschrig H, Perlov A Y, and   
Oppeneer P M 1998 Phys.\ Rev.~B~{\bf 58}, 293    
   
\bibitem{spinpaper} B\"unemann J 2001 J. Phys. Cond. Matt. {\bf 13} 5327   
 
\bibitem{florianbuch} Gebhard F 1997 The Mott   
Metal-Insulator Transition (Berlin: Springer)   
  
\bibitem{Mahan} Mahan G D 1990 Many-Particle Physics 2nd ed.  
(New York: Plenum Press)  
  
\bibitem{GebPRB90} Gebhard F 1990 Phys. Rev. B {\bf 41}, 9452  
 
\bibitem{wbp} B\"unemann J, in preparation  
  
\end{references}
\end{document}